\documentclass[]{krakproc}
\usepackage{graphicx}
\usepackage{bm}
\newcommand{\EQ}{\begin{equation}}
\newcommand{\EN}{\end{equation}}
\newcommand{\EQA}{\begin{eqnarray}}
\newcommand{\ENA}{\end{eqnarray}}

\newcommand{\Eq}[1]{equation~(\ref{#1})}

\newcommand{\Sec}[1]{\S\ref{#1}}

\newcommand{\Fig}[1]{Fig.~\ref{#1}}
\newcommand{\FFig}[1]{Figure~\ref{#1}}

\newcommand{\bra}[1]{\langle #1\rangle}

\newcommand{\meanemf}{\overline{\mbox{\boldmath ${\cal E}$}}{}}{}

\newcommand{\meanBB}{\overline{\bm{B}}}
\newcommand{\meanJJ}{\overline{\bm{J}}}
\newcommand{\meanUU}{\overline{\bm{U}}}
\newcommand{\meanWW}{\overline{\bm{W}}}

{}
{}
{}
{}
{}
{}
{}

\newcommand{\UU}{{\bm{U}}}

\newcommand{\uu}{{\bm{u}}}
\newcommand{\BB}{{\bm{B}}}

\newcommand{\jj}{{\bm{j}}}

\newcommand{\bb}{{\bm{b}}}

\newcommand{\nn}{\mbox{\boldmath $n$} {}}

\newcommand{\nab}{\mbox{\boldmath $\nabla$} {}}
\newcommand{\OO}{\mbox{\boldmath $\Omega$} {}}
\newcommand{\oo}{\mbox{\boldmath $\omega$} {}}

\newcommand{\dd}{{\rm d} {}}

\def\Ma{\mbox{\rm Ma}}

\def\Rey{\mbox{\rm Re}}

\def\onethird{{\textstyle{1\over3}}}

\newcommand{\yr}{\,{\rm yr}}

\newcommand{\yapj}[3]{ #1, {ApJ,} {#2}, #3}

\newcommand{\yan}[3]{ #1, {AN,} {#2}, #3}

\newcommand{\yana}[3]{ #1, {A\&A,} {#2}, #3}

\newcommand{\yjfm}[3]{ #1, {JFM,} {#2}, #3}
\newcommand{\ypf}[3]{ #1, {Phys. Fluids,} {#2}, #3}

\newcommand{\yjetp}[3]{ #1, {Sov. Phys. JETP,} {#2}, #3}

\newcommand{\yannr}[3]{ #1, {ARA\&A,} {#2}, #3}

\newcommand{\yprl}[3]{ #1, {PRL,} {#2}, #3}

\newcommand{\ymn}[3]{ #1, {MNRAS,} {#2}, #3}

\newcommand{\ypre}[3]{ #1, {PRE,} {#2}, #3}

\newcommand{\yjour}[4]{ #1, {#2}, {#3}, #4}

\newcommand{\ybook}[3]{ #1, {#2} (#3)}

\begin{document}

\title{Nonhelical turbulent dynamos: shocks and shear}
\author{A.\ Brandenburg$^{1,2}$, N.\ E.\ L.\ Haugen$^{3,4}$, \& A.\ J.\ Mee$^5$}
\institute{
$^1$Isaac Newton Institute for Mathematical Sciences,
20 Clarkson Road, Cambridge CB3 0EH, UK\\
$^2$Nordita, Blegdamsvej 17, DK-2100 Copenhagen \O, Denmark\\
$^3$DAMTP, University of Cambridge,
Wilberforce Road, Cambridge CB3 0WA, UK\\
$^4$Department of Physics, NTNU,
H{\o}yskoleringen 5, N-7034 Trondheim, Norway\\
$^5$School of Mathematics and Statistics, University of Newcastle
upon Tyne, NE1 7RU, UK}
\markboth{A.\ Brandenburg, N.\ E.\ L.\ Haugen, \& A.\ J.\ Mee}
{Overview of dynamos in stars and galaxies}

\maketitle

\begin{abstract}
Small scale turbulent dynamo action in compressible transonic turbulence
is discussed.
It is shown that the critical value of the magnetic Reynolds number
displays a bimodal behavior and changes from a typical value of 35 for
small Mach numbers to about 80 for larger Mach numbers.
The transition between the two regimes is relatively sharp.
The direct simulations are then compared with simulations where
shocks are captured using a shock viscosity that becomes large at
locations where there are shocks.
In the presence of shear it is shown that large scale dynamo action
is possible.
\end{abstract}

\section{Introduction}

There are two main classes of dynamos, small scale and large scale dynamos.
This distinction is not completely unambiguous because in the presence
of shear any small scale field will attain a large scale component in
the direction of the shear.
Therefore, the class of small scale dynamos is here confined to the case of
nonhelical, isotropic and homogeneous turbulent flows (\Sec{SSdynamos}).
With shear, however, even nonhelical dynamos can produce large scale
fields (\Sec{LSdynamos}).

Members of the large scale dynamo class include turbulent flows with
sufficient amounts of helicity and/or shear such that a large scale
field is generated.
Here, a large scale field is the component that survives after averaging.
The averaging procedure has to be defined appropriately and depends on
circumstances and also on what kind of field is generated.
In the presence of a shear flow, $\UU_0(x,z)=(0,U_0,0)$, for example,
a useful average (denoted by an overbar) will be
\EQ
\meanBB={1\over L_y}\int_0^{L_x}\BB\,\dd y.
\label{AzimthalAverage}
\EN
This definition will obviously preclude the study of nonaxisymmetric fields.
In the case of helical turbulence, we have essentially an $\alpha$ effect,
i.e.\ the turbulent electromotive force has a field-aligned component and
$\overline{\uu\times\bb}\cdot\meanBB\neq0$.
Here, $\bb=\BB-\meanBB$ is the fluctuating magnetic field and
$\uu=\UU-\meanUU$ is the fluctuating velocity.
In unbounded space, for example, the mean field can have any orientation.
Even in a triply-periodic domain there are still three ultimate field
configurations, corresponding to Beltrami waves with variation in any
of the three coordinate directions.
The appropriate mean field is then best defined as a two-dimensional
average over the other two coordinate directions.

Ideally, when defining averages we want to conform with the
Reynolds rules (e.g.\ Krause \& R\"adler 1980).
In particular, we want to make sure that the average of an average gives
the same average (which is not the case for a running mean), and that
average and derivative operators commute (not the case for averages over
non-periodic directions, including time averages).
We also want the average of a product of fluctuating and mean quantities
to vanish (which is not the case for spectral filtering).
Therefore, for many practical purposes, \Eq{AzimthalAverage} is the
preferred choice, avoiding any of the aforementioned problems.

A more formal characteristics of a {\it successful} large scale dynamo
is therefore one where the ratio
\EQ
q\equiv\bra{\meanBB^2}/\bra{\BB^2}\gg R_{\rm m}^{-1}.
\EN
Here, angular brackets denote volume averages.
Values of $q$ around 0.7 are typical under ideal conditions
(see \Sec{LSdynamos}).
Spiral galaxies tend to have $q\ge0.2$ (e.g.\ Beck et al.\ 1996).
For the sun the value of $q$ is unclear, but we would still classify it
as a large scale dynamo even if $q$ was as small as 0.01, say.

In the following we give a brief overview of recent progress in the
fields of small scale and large scale dynamos.
In this review we focus on numerical results.

\section{Small scale dynamos}
\label{SSdynamos}

Small scale dynamos are generally much harder to excite than large
scale dynamos.
In fact, for unit magnetic Prandtl numbers
($\mbox{Pr}_{\rm M}\equiv\nu/\eta=1$) the critical value of the
magnetic Reynolds number, $R_{\rm m}=u_{\rm rms}/(\eta k_{\rm f})$,
is 35 for the small dynamo (e.g., Haugen et al.\ 2004a) and only 1.2
for fully helical dynamos (Brandenburg 2001).
Here we should emphasize that there is an advantage in defining
$R_{\rm m}$ with respect to the wavenumber instead of the forcing
scale (which would make $R_{\rm m}$ larger by $2\pi$) or even the
scale of the box (which would make $R_{\rm m}$ larger by the scale
separation ratio).
The advantage is that with our definition (which is actually quite
common in the forced turbulence community) the value of $R_{\rm m}$
can be regarded as a reasonable approximation to $\eta_{\rm t}/\eta$,
where $\eta_{\rm t}$ is the turbulent (effective) magnetic diffusivity
and $\eta$ is the microscopic value.

In the following we summarize what is now known about the energy
spectra in the linear and nonlinear regimes and what happens in the
presence of shocks.

\subsection{Kazantsev and Kolmogorov spectra}

\begin{figure}[t!]\centering
\includegraphics[width=.50\textwidth]{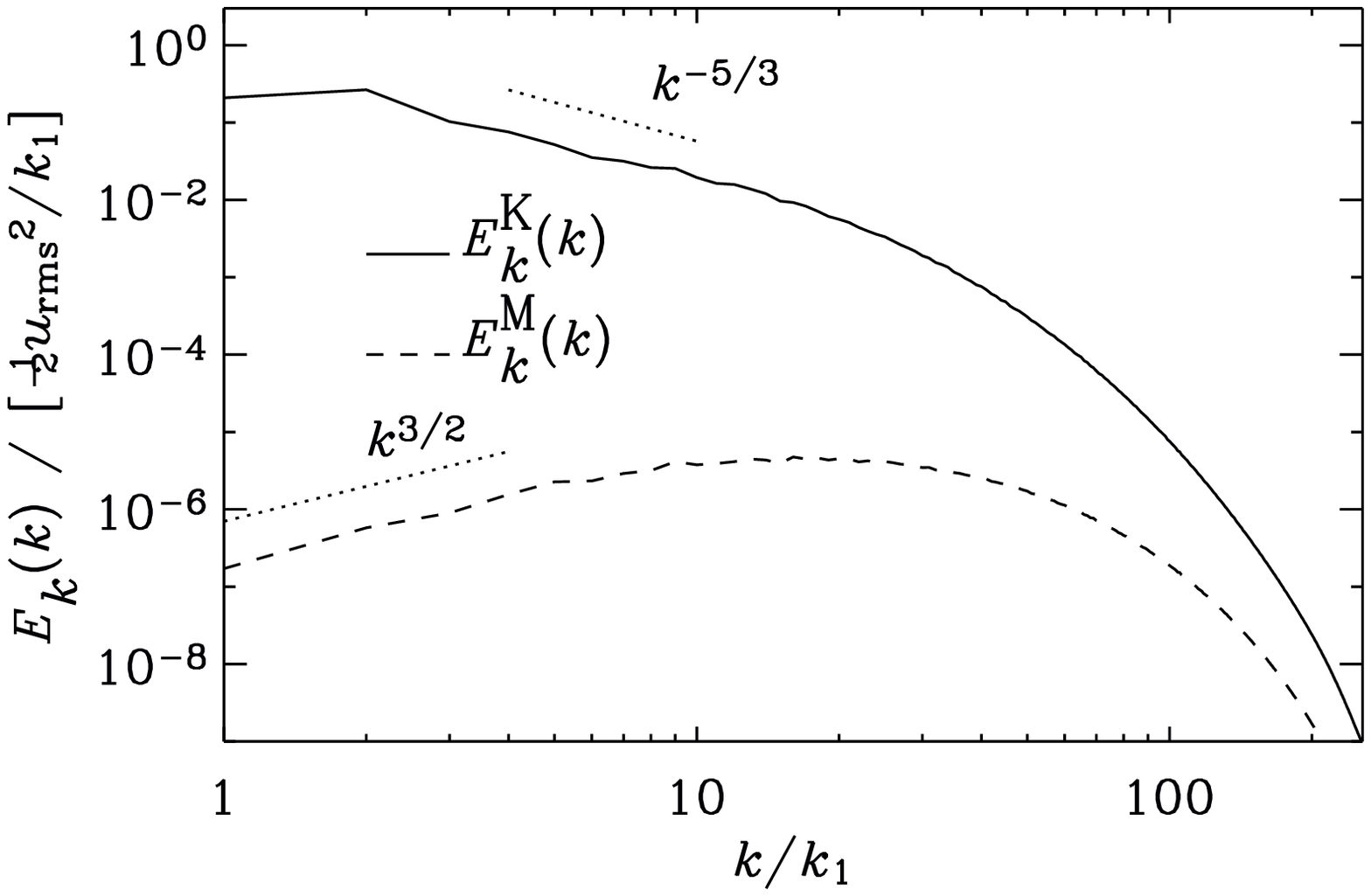}
\includegraphics[width=.45\textwidth]{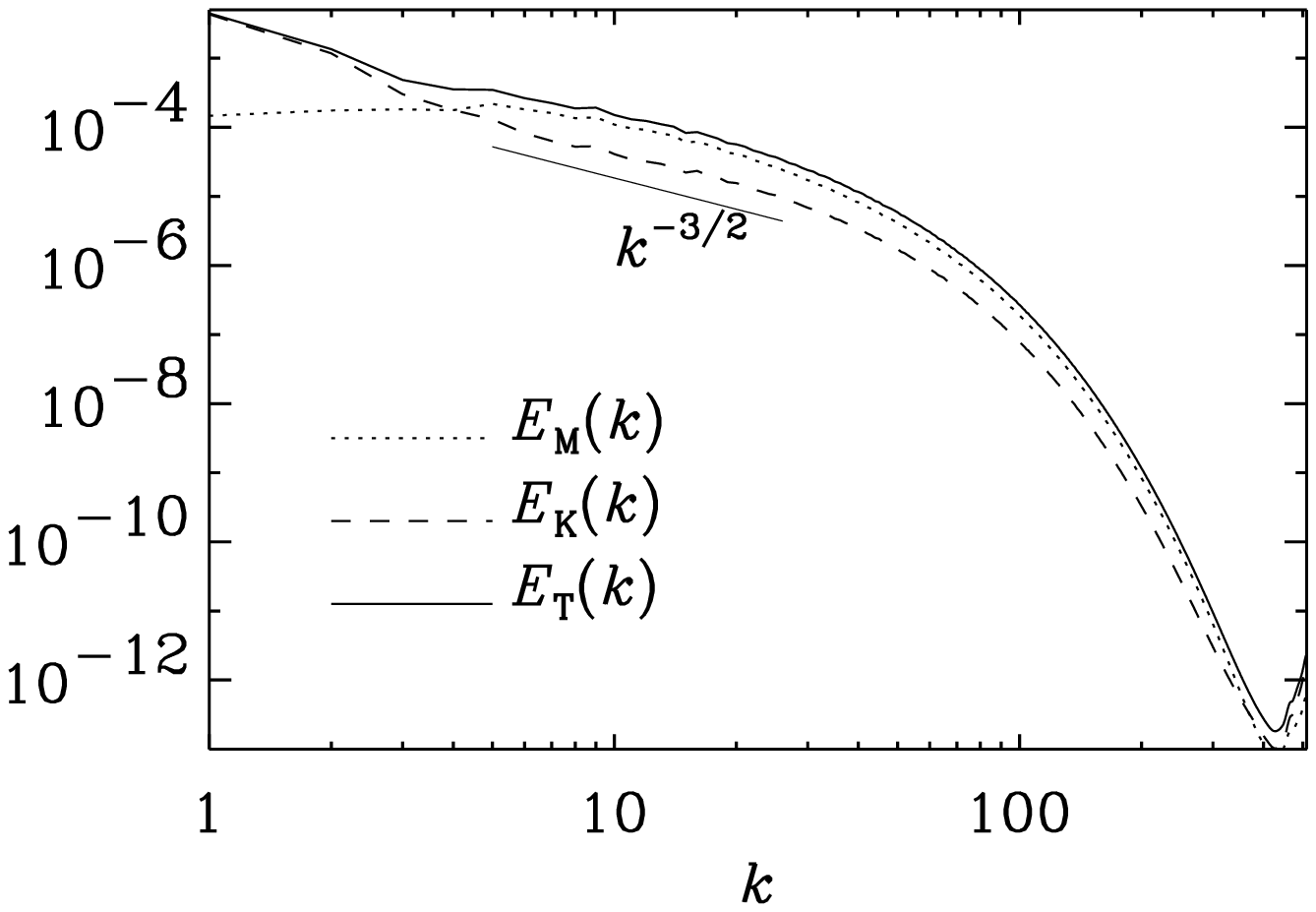}
\caption{
Left: early spectra of kinetic and magnetic energy,
normalized by $\frac{1}{2}u_{\rm rms}^2/k_1$,
during the kinematic stage of run D2.
[Adapted from Haugen et al.\ (2004a).]
Right: magnetic, kinetic and total energy spectra.
$1024^3$ meshpoints.
The Reynolds number is $u_{\rm rms}/(\nu k_{\rm f})\approx960$.
[Adapted from Haugen et al.\ (2003).]
}\label{plot_poweru_early}\end{figure}

In the kinematic regime, the magnetic field is still weak and so the
velocity field is like in the nonmagnetic case, with the usual Kolmogorov
spectrum followed by a dissipative subrange.
Due to an extended bottleneck effect\footnote{
For details regarding the bottleneck effect see the recent paper
by Dobler et al.\ (2003), where the difference between the fully
three-dimensional spectra and the one-dimensional spectra available from
wind tunnel experiments is explained.}, only marginally
indicated; see \Fig{plot_poweru_early} for a simulation by
Haugen et al.\ (2004a) where $\mbox{Pr}_{\rm M}=1$ and $R_{\rm m}=600$.
During the kinematic stage, the magnetic field shows a clear $k^{3/2}$
spectrum that is characteristic of the Kazantsev (1968) spectrum that
was originally only expected in the large magnetic Prandtl number limit,
i.e.\ for $\mbox{Pr}_{\rm M}\equiv\nu/\eta\gg1$.
During the kinematic phase the spectral magnetic energy grows at
all wavenumbers exponentially in time and the spectrum remains
shape-invariant.

As the magnetic energy increases, the spectrum arranges itself underneath
an envelope given by the original kinetic energy spectrum.
During this process, the kinetic energy decreases by a certain amount and,
above a certain wavenumber, the field can be in super-equipartition with
the velocity; see the right hand panel of
\Fig{plot_poweru_early} for a high resolution run with $1024^3$
meshpoints (Haugen et al.\ 2003).
These spectra are, as usual, integrated
over shells in $k$ space and normalized such that
$\int E_{\rm K}\dd k={1\over2}\bra{\uu^2}$ and
$\int E_{\rm M}\dd k={1\over2}\bra{\BB^2}/\mu_0$.
The magnetic energy displays a nearly flat spectrum in the range
$1\leq k\leq5$, peaks at $k\approx5$,
and begins to exhibit an inertial range in $8\leq k\leq25$, followed by
a dissipative subrange over one decade.
In the inertial range $E_{\rm M}(k)/E_{\rm K}(k)$ is about 2.5.

\begin{figure}[t!]\centering\includegraphics[width=.5\textwidth]
{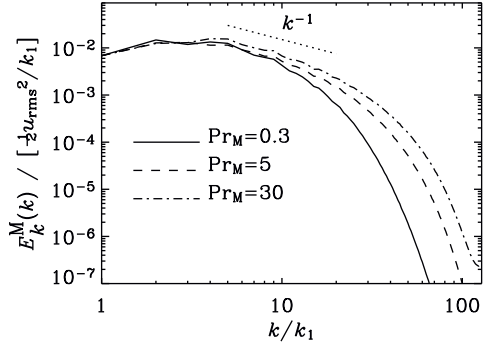}\caption{
Magnetic energy spectra for runs with magnetic Prandtl numbers ranging
from 0.3 to 30.
[Adapted from Haugen et al.\ (2004a).]
}\label{prandtl256}\end{figure}

At larger magnetic Prandtl numbers one begins to see the possible
emergence of a $k^{-1}$ tail in the magnetic energy spectrum; see
\Fig{prandtl256}.
The $k^{-1}$ tail has recently been found in large magnetic Prandtl number
simulations with an imposed magnetic field (Cho et al.\ 2002).
The $k^{-1}$ spectrum has its roots in early work by Batchelor (1959)
for a passive scalar and Moffatt (1963) for the magnetic case.
In the low magnetic Prandtl number case, evidence is accumulating that
the critical magnetic Reynolds number continues to become larger
(Schekochihin et al.\ 2004, Boldyrev \& Cattaneo 2004, Haugen et al.\ 2004).

\begin{figure}[t!]\centering
\includegraphics[width=.40\textwidth]{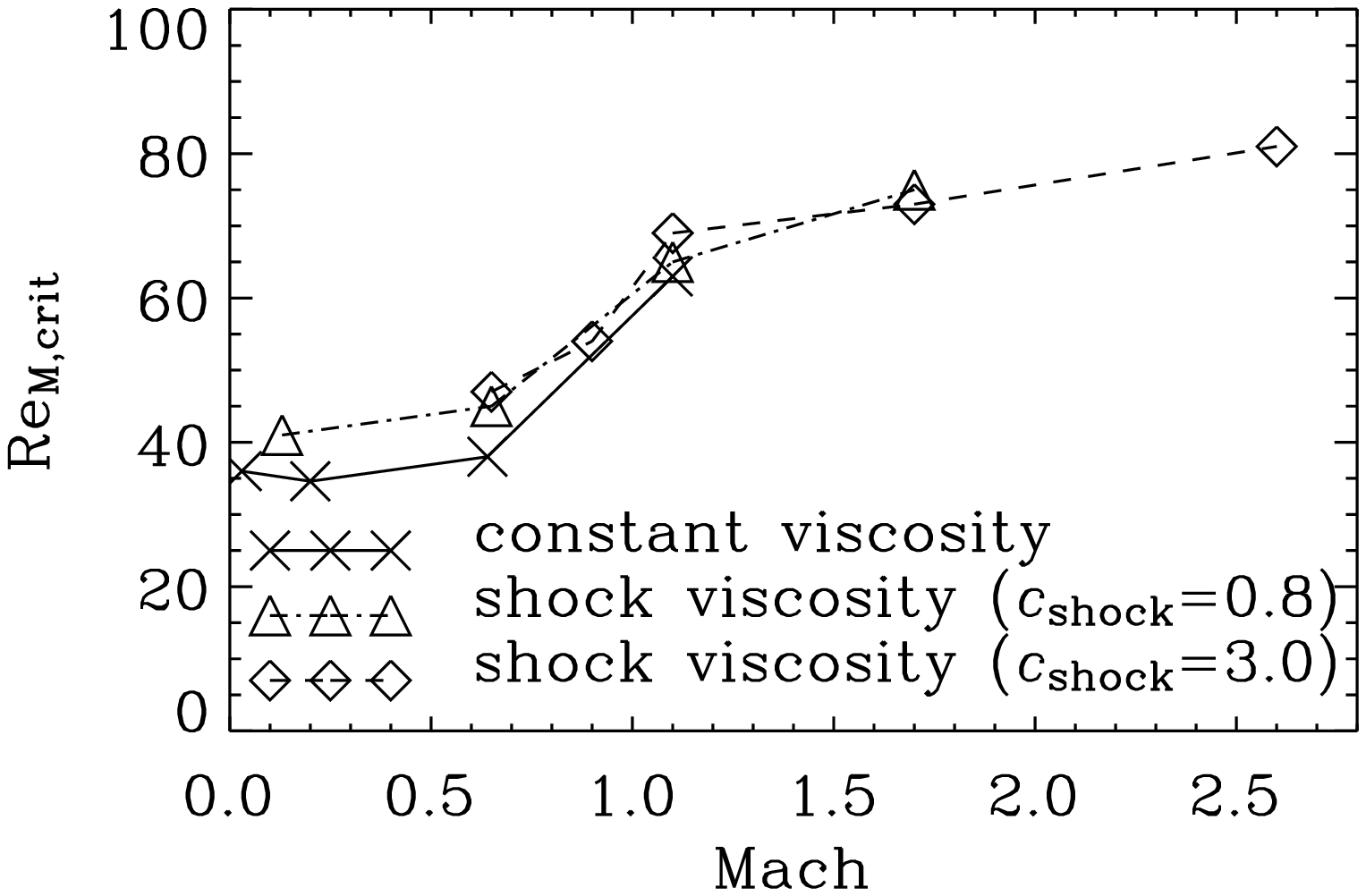}
\includegraphics[width=.40\textwidth]{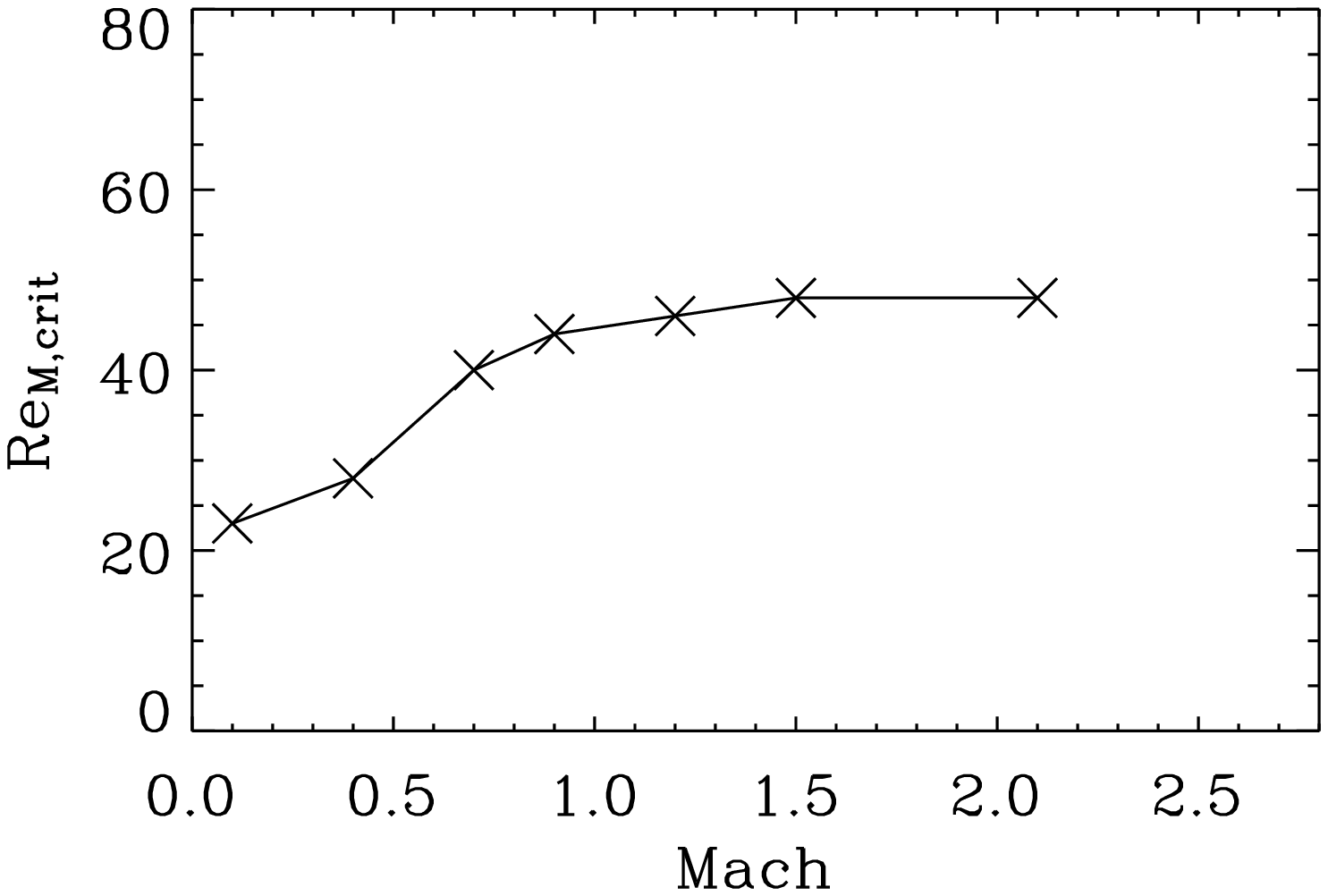}\caption{
Critical magnetic Reynolds number $\Rey_{\rm M,crit}$ as a function of $\Ma$
for simulations with $\mbox{Pr}_{\rm M}=1$ (left)
and $\mbox{Pr}_{\rm M}=5$ (right).
Note that $\Rey_{\rm M,crit}$ depends strongly on Mach number for
$\Ma\approx1$.
The simulations with shock-capturing viscosity give approximately
the correct growth rates. The simulations that provide these data points
have resolutions ranging from $64^3$ to $512^3$ mesh points.
[Adapted from Haugen et al.\ (2004b).]
}\label{Mach_dep_Rmcrit}\end{figure}

\begin{figure}\centering
\includegraphics[width=.48\textwidth]{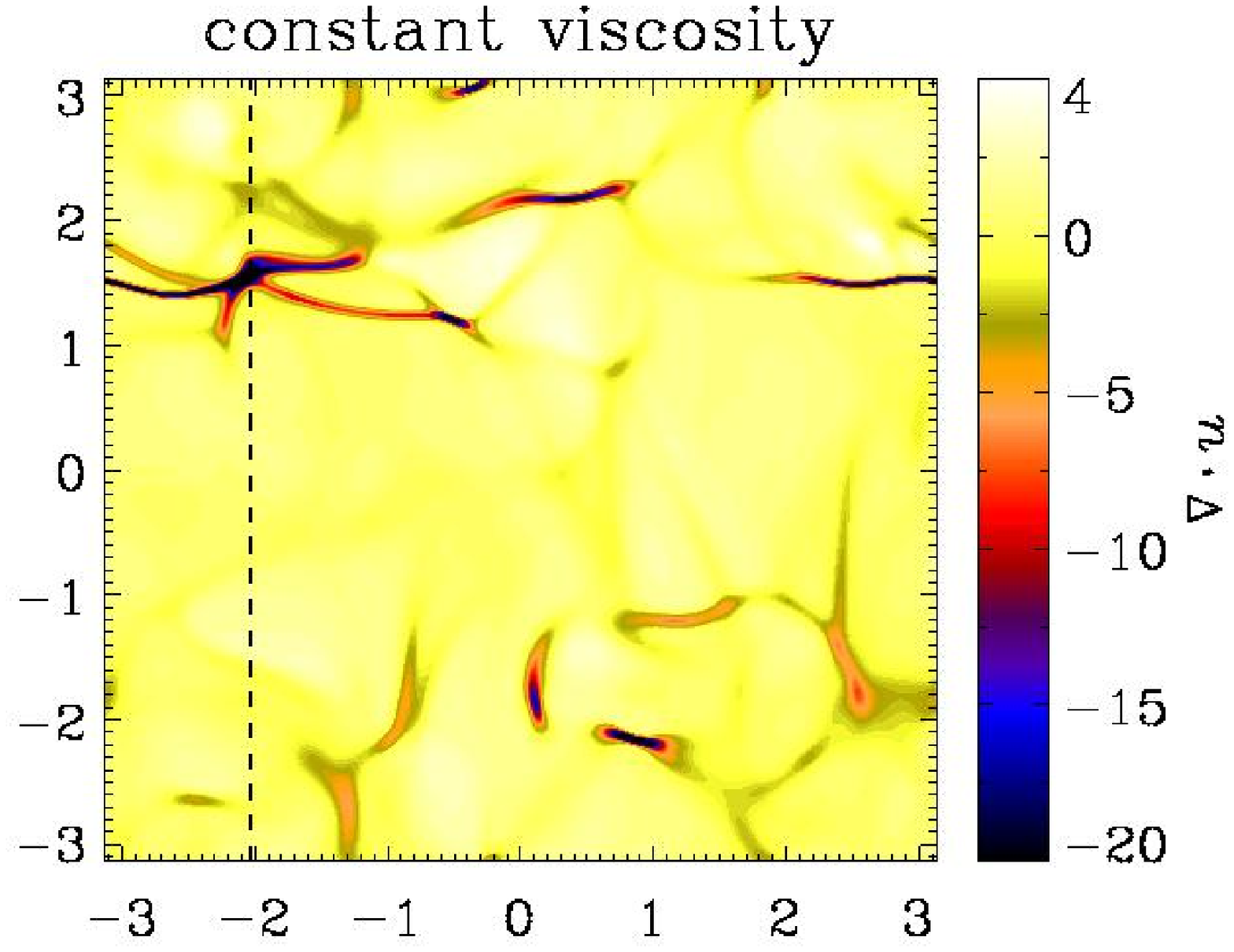}
\includegraphics[width=.32\textwidth]{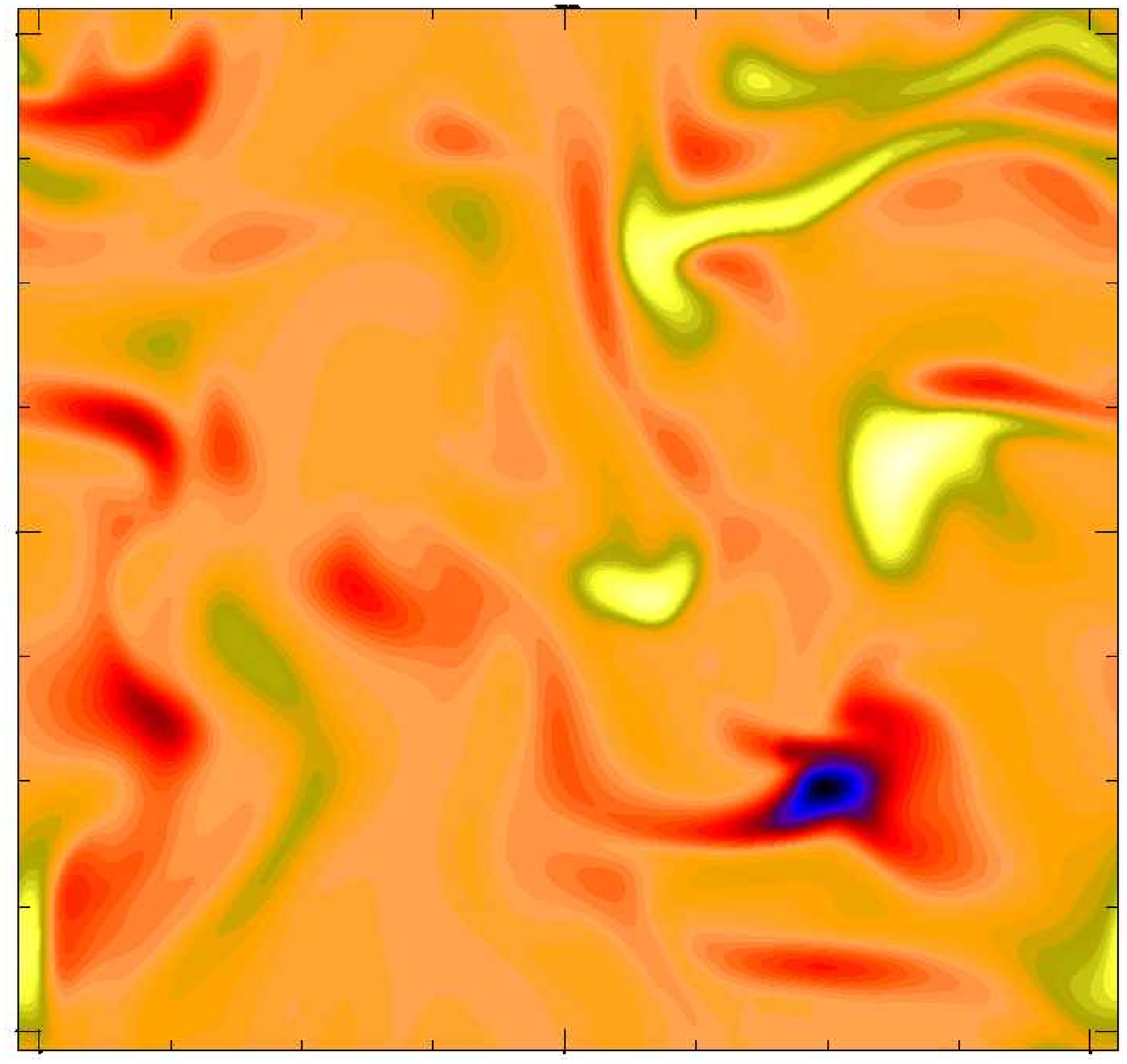}\caption{
Grey (or color) scale representation of $\nab\cdot\uu$ (left)
and $B_z$ (right) in an $xy$ cross-section through $z=0$
for $\Ma=1.1$ using constant viscosity (Run~3a of Haugen et al.\ 2004b
with $512^3$ meshpoints).
[Adapted from Haugen et al.\ (2004b).]
}\label{vis_bb}\end{figure}

\subsection{Do shocks kill dynamos?}

In the interstellar medium the flows are generally supersonic and
in places even highly supersonic.
Naively, one could think of highly supersonic turbulence as being
nearly irrotational, in which case the dynamo should actually become
much more efficient and the growth rate should
increase with Mach number (Kazantsev et al.\ 1985).
This is now known not to be the case: in a mixture of irrotational
($\oo\equiv\nab\times\uu=0$) and solenoidal ($\nab\cdot\uu=0$) flows
the critical values of the magnetic Reynolds number for the onset
of dynamo action does actually increase as a function of the ratio
$\sigma\equiv\bra{(\nab\cdot\uu)^2}/\bra{\oo^2}$
(Rogachevskii \& Kleeorin 1997).
However, the ratio $\sigma$ is found to stay finite in the limit
of large Mach numbers (Porter et al.\ 1998, Padoan \& Nordlund 1999).
This explains the recent finding that minimum magnetic Reynolds number
for dynamo action displays a bimodal behavior as a function of the
Mach number $\mbox{Ma}=u_{\rm rms}/c_{\rm s}$ (Haugen et al.\ 2004b).
In fact, for $\mbox{Pr}_{\rm M}=1$, they find
$R_{\rm m,crit}\approx35$ for $\mbox{Ma}\ll1$ and
$R_{\rm m,crit}\approx80$ for $\mbox{Ma}\gg1$.
For $\mbox{Pr}_{\rm M}=5$, the critical values are a bit lower
(25 and 50 respectively); see \Fig{Mach_dep_Rmcrit}.
Indeed, in these simulations the ratio $\sigma$ is always about 1/4.
Visualizations of the magnetic field show that the effects of shocks
is rather weak; see \Fig{vis_bb}, where we compare both cases.

\begin{figure}\centering\includegraphics[width=0.6\textwidth]
{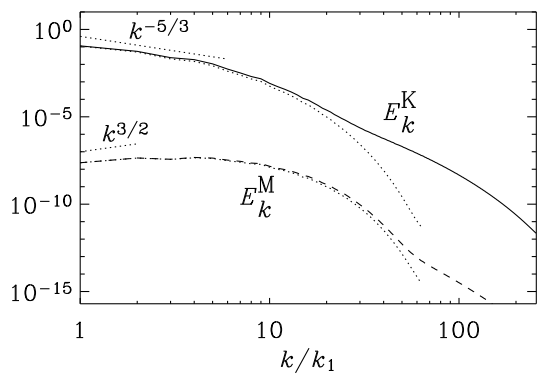}\caption{
Energy spectra for Runs~1a and 1c of Haugen et al.\ (2004b).
The dotted lines give the result using shock-capturing viscosity.
[Adapted from Haugen et al.\ (2004b).]
}\label{spectra}\end{figure}

It is worth noting that the use of a shock capturing viscosity
({\it not} used in the calculations presented in \Fig{vis_bb}) seems to
reproduce the critical values of the magnetic Reynolds number rather well;
see the left hand panel of \Fig{Mach_dep_Rmcrit}.
In \Fig{spectra} we compare kinetic and magnetic energy spectra for
the constant and shock-capturing viscosity solutions.
The two show excellent agreement at low wavenumbers.
At high wavenumbers there are major differences.
The direct simulations show an extended diffusion range which is
rather from the diffusive subrange in incompressible simulations
(e.g.\ Kaneda et al.\ 2003).
The large extent of this diffusive subrange in supersonic turbulence
is obviously the reason for the tremendously high resolution needed
in the direct simulations.
Fortunately, as far as dynamo action is concerned, this extended subrange
does not need to be fully resolved and it suffices to just cut it off with
a shock-capturing viscosity.

 \section{Large scale dynamos}
\label{LSdynamos}

There are some major uncertainties in what exactly are the relevant
large scale dynamo mechanisms in galaxies and stars.
Simulations have enabled us to make close contact between simulations
and theory.
We are therefore beginning to see some significant progress in that
many of the uncertainties intrinsic to mean field dynamo theory
can now be eliminated.
However, the anticipated agreement between theory and simulations does
not yet extend to the more realistic cases with strong inhomogeneities
and anisotropies, for example where the turbulence is driven by convection,
supernovae, or by the magneto-rotational instability.

Crucial to improving the agreement between theory and simulations
has been the realization that the dominant nonlinear feedback comes
from the current helicity term.
This term can produce a ``magnetic $\alpha$ effect'' (which is
$\alpha_{\rm M}=\onethird\tau\overline{\jj\cdot\bb}/\rho_0$ in the
isotropic case) -- even if there is no kinetic $\alpha$ effect.
The latter is $\alpha_{\rm K}=-\onethird\tau\overline{\oo\cdot\uu}$
in the isotropic case,
where $\oo=\nab\times\uu$ is the vorticity, $\jj=\nab\times\bb/\mu_0$
the current density, and $\rho_0$ the mean density.
An example where $\alpha_{\rm M}$ can be generated -- even with
$\alpha_{\rm K}=0$ -- is the shear-current effect of
Rogachevskii \& Kleeorin (2003, 2004).

\begin{figure}[t!]\centering\includegraphics[width=.55\textwidth]
{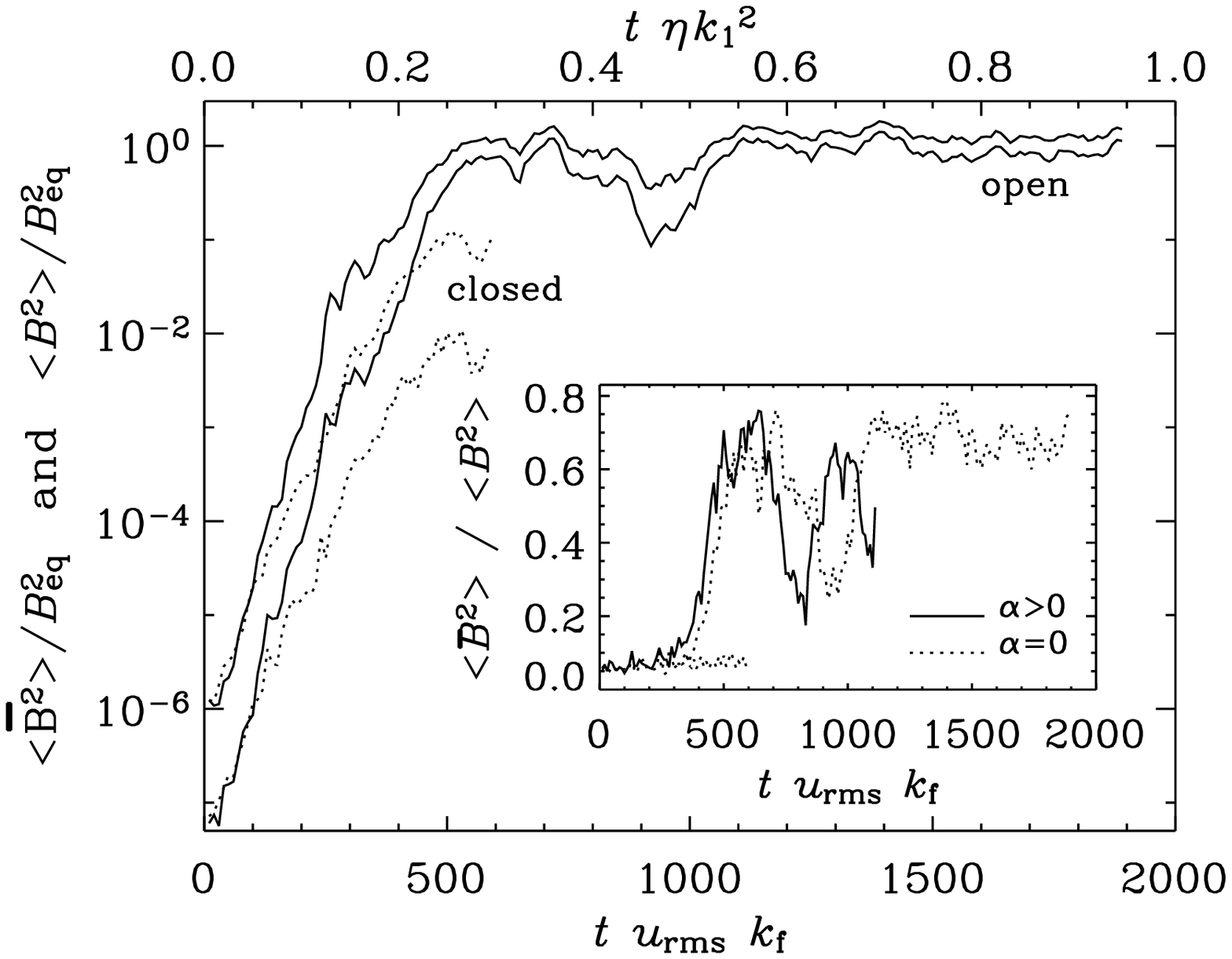}\caption{
Evolution of the energies of the total field $\bra{\BB^2}$ and of
the mean field $\bra{\meanBB^2}$, in units of $B_{\rm eq}^2$,
for runs with non-helical forcing
and open or closed boundaries; see the solid and dotted lines, respectively.
The inset shows a comparison of the ratio $\bra{\meanBB^2}/\bra{\BB^2}$
for nonhelical ($\alpha=0$) and helical ($\alpha>0$) runs.
For the nonhelical case the run with closed boundaries is also
shown (dotted line near $\bra{\meanBB^2}/\bra{\BB^2}\approx0.07$).
Note that saturation of the large scale field occurs on a
dynamical time scale; the resistive time scale is given on the
upper abscissa.
}\label{pmean_comp}\end{figure}

Recently, quantitative comparisons between theory and
simulations have shown that whatever mean electromotive force
($\meanemf\equiv\overline{\uu\times\bb}$) is produced by the mean field
dynamo, this produces magnetic helicity in the large scale field.
Because of magnetic helicity conservation, a corresponding negative
contribution in the magnetic helicity of the small scale magnetic field
must be generated.
This results in the production of a small scale current helicity
which enters in the mean field equations as a magnetic alpha effect.
To satisfy magnetic helicity conservation at all times, the small
scale magnetic helicity equation (or equivalently the evolution
equation for $\alpha_{\rm M}$) has to be solved simultaneously
with the mean field equations.
This reproduces quantitatively the resistively slow saturation of
helically forced dynamos in a periodic box (Field \& Blackman 2002,
Blackman \& Brandenburg 2002, Subramanian 2002).
The explicitly time-dependent evolution equation for the magnetic
$\alpha$ effect was first derived by Kleeorin \& Ruzmaikin (1982).
Early work by Ruzmaikin (1981) focused attention on the possibility
of chaotic dynamics introduced by this effect (see also subsequent
work by Schmalz \& Stix 1991 and Covas et al.\ 1997),
and the first connection between dynamical and catastrophic quenching
was made by Kleeorin et al.\ (1995).

In order to avoid too much repetition with recent reviews on this subject
(e.g., Brandenburg et al.\ 2002) we discuss here only in a few words
the main consequences of the dynamical quenching model.
It reproduces quantitatively the resistively slow saturation phase of
an $\alpha^2$ dynamo in a periodic box (Field \& Blackman 2002).
Secondly, it reproduces reasonably accurately the saturation amplitude
and the cycle period of $\alpha\Omega$ dynamos with a sinusoidal shear
profile in a periodic box.

In the case of galaxies the dynamical time scale is $10^7\yr$, which
puts rather stringent constraints if one wants to explain microgauss
field strengths in very young galaxies that are $10^9\yr$ old.
It is likely that a successful dynamo has to have magnetic helicity
fluxes that allow the dynamo to get rid of small scale magnetic helicity
to allow for a rapid build-up of a large scale field that tends to
have magnetic helicity of the opposite sign (Blackman \& Field 2000,
Kleeorin et al.\ 2000).
The currently most convincing example where the presence of boundaries
has been found to be important is in connection with turbulent dynamos
that work mainly in the presence of shear (Brandenburg et al.\ 2005).
Here, the main effect that is thought to be responsible is the
so-called shear--current effect where the electromotive force has
a component proportional to $\meanWW\times\meanJJ$. Here, $\meanWW$ is
the vorticity of the mean flow and $\meanJJ$ is the mean current density.
This effect is technically related to the $\OO\times\meanJJ$ effect
(e.g.\ Krause \& R\"adler 1980), which is obviously quite distinct from the
famous $\alpha$ effect.
Another possibility is the Vishniac \& Cho (2001) mechanism.
However, it has not yet been possible to verify any of them explicitly.
Preliminary mean field calculations suggest, however, that the $\meanWW\times\meanJJ$
effect produces qualitatively favorable agreement with the direct simulations.

\FFig{pmean_comp} shows that in the presence of closed (i.e.\ perfectly
conducting) boundaries, magnetic energy both of the total field
and of the mean field saturates at a much lower level than with open
boundaries ($\nn\times\BB=0$).
The inset shows the ratio between the energies contained in the mean field
to that in the total field.
Note that this ratio stays small when the boundaries are closed, but it increases to
fairly large values (around 0.7) when the boundaries are open.

\section{Conclusions}

In many astrophysical bodies some type of large scale dynamo is likely
to operate.
This dynamo may operate with nonhelical turbulence and shear alone,
i.e.\ without $\alpha$ effect.
Small scale dynamos, on the other hand, are an extreme
case which requires that there is no shear and no helicity, for example.
Both types of dynamos are vulnerable in their own ways: a large scale
dynamo requires magnetic and current helicity fluxes in order to be successful,
while small scale dynamos may require magnetic Prandtl numbers that are
not too small.
However, as we have shown in the present paper, small scale dynamos
still work in the compressible regime.
In fact, it now seems that, once the Mach number exceeds unity, its
onset becomes independent of the Mach number.
This hypothesis has only been tested for small Mach numbers, so it would
be useful to extent these studies to larger values.
However, as we have also been able to show,
the use of shock-capturing viscosities seems to be a reasonably accurate
approximation for this purpose.

\section*{Acknowledgements}

The Danish Center for Scientific Computing is acknowledged
for granting time on the Linux cluster in Odense (Horseshoe).

\end{document}